\documentclass[showpacs,aps,prd,preprint,nofootinbib,showkeys,unsortedaddress,raggedbottom]{revtex4-1}
\pdfoutput=1

\usepackage{bm}
\usepackage{amsmath}
\usepackage{graphicx}
\usepackage{subfigure}
\usepackage[usenames,dvipsnames]{color}
\definecolor{darkblue}{RGB}{0,0,196}
\usepackage[colorlinks=true,linkcolor=darkblue,citecolor=darkblue,urlcolor=darkblue]{hyperref}

\usepackage{setspace}
\usepackage{footmisc}
\usepackage[makeroom]{cancel}

\newcommand{\pres}{{\cal P}}
\newcommand{\preseq}{\pres_{\rm eq}}
\newcommand{\taueq}{\tau_{\rm eq}}

\def\be{\begin{equation}}
\def\ee{\end{equation}}
\def\ba{\begin{eqnarray}}
\def\ea{\end{eqnarray}}

\begin{document}

\title{Including off-diagonal anisotropies in anisotropic hydrodynamics}

\author{Mohammad Nopoush} 



\author{Michael Strickland} 

\affiliation{Department of Physics, Kent State University, Kent, OH 44242 United States}

\begin{abstract}
In this paper we present a method for efficiently including the effects of off-diagonal local rest frame momentum anisotropies in leading-order anisotropic hydrodynamics.  The method relies on diagonalization of the space-like block of the anisotropy tensor and allows one to reduce the necessary moments of the distribution function in the off-diagonal case to a linear combination of diagonal-anisotropy integrals.  Once reduced to diagonal-anisotropy integrals, the results can be computed efficiently using techniques described previously in the literature.  We present a general framework for how to accomplish this and provide examples for off-diagonal anisotropy moments entering into the energy-momentum tensor and viscous update equations which emerge when performing anisotropic pressure matching.
\end{abstract}

\date{\today}



\maketitle

\section{Introduction}

Ultra-relativistic heavy ion collision (URHIC) experiments, e.g. RHIC at BNL and LHC at  CERN, aim to study the dynamics and properties of matter at extremely high-energy density. In these experiments, matter is heated to temperatures exceeding the QCD pseudo-critical temperature, $T_{\rm pc} \simeq 155$ MeV, using ultra-relativistic collisions among heavy nuclei, protons, deuterons, etc. The strongly interacting droplet of matter produced during high-energy and high-multiplicity URHICs is called the quark-gluon plasma (QGP).  In high-multiplicity events, the QGP demonstrates strong collective behavior during evolution from hydrodynamization ($\tau \sim 1$ fm/c) to hadronic freeze out ($\tau \sim 10$ fm/c). During this time period it has been found that relativistic fluid dynamics formalisms can effectively describe the evolution of the system and one finds that information about initial state geometry of the target (average eccentricity and fluctuations) is reflected in final state observables, e.g. the azimuthal dependence of hadron production. In other words, one can track the correlations between the eccentricity of the initial state's geometry and the flow harmonics observed in the final state hadron spectra using dissipative hydrodynamics.  The success of relativistic dissipative hydrodynamics \cite{Heinz:2013th,Jeon:2016uym,Romatschke:2017ejr,Alqahtani:2017mhy} has inspired theoreticians to make the underlying formalisms more complete and robust with respect to large deviations from isotropic thermal equilibrium using standard fixed-order viscous hydrodynamics (vHydro) treatments \cite{Muller:1967zza,Israel:1976tn,Israel:1979wp,Muronga:2001zk,Muronga:2003ta,Muronga:2004sf,Heinz:2005bw,Baier:2006um,Romatschke:2007mq,Baier:2007ix,Dusling:2007gi,Luzum:2008cw,Song:2008hj,Heinz:2009xj,Schenke:2010rr,Schenke:2011tv,Bozek:2011wa,Niemi:2011ix,Denicol:2011fa,Niemi:2012ry,Bozek:2012qs,Denicol:2012cn,Denicol:2012es,Jaiswal:2013npa,Jaiswal:2013vta,Denicol:2014vaa,Denicol:2014mca,Jaiswal:2014isa} and resummed anisotropic hydrodynamics (aHydro) treatments \cite{Florkowski:2010cf,Martinez:2010sc,Ryblewski:2010ch,Florkowski:2011jg,Martinez:2012tu,Ryblewski:2012rr,Bazow:2013ifa,Tinti:2013vba,Nopoush:2014pfa,Florkowski:2014bba,Tinti:2015xwa,Bazow:2015cha,Bazow:2015zca,Nopoush:2015yga,Alqahtani:2015qja,Molnar:2016vvu,Molnar:2016gwq,Bluhm:2015raa,Bluhm:2015bzi,Alqahtani:2017jwl,Alqahtani:2017tnq,Alqahtani:2017mhy,Almaalol:2018gjh}.  

The introduction of the aHydro formalism was driven by the fact that, due to the strong early-stage longitudinal expansion of the QGP, one finds large momentum-space anisotropy in the local rest frame (LRF) of the QGP which persists for many fm/c. The magnitude of the momentum-space anisotropy has cast some doubt on the quantitative accuracy of standard vHydro which assumes that one can linearize around isotropic equilibrium.  aHydro is a non-equilibrium hydrodynamics model which takes into account the strong momentum-space anisotropy of the QGP at leading order and in doing so resums an infinite number of terms in inverse Reynolds number \cite{Strickland:2017kux}.  In contrast to standard vHydro, aHydro is based on Taylor expansion about an anisotropic distribution function instead of an isotropic one. This allows one to capture the dominant anisotropic contributions to the distribution function in the leading order term, thereby guaranteeing positivity of the one-particle distribution at all space-time points at leading-order.  aHydro and vHydro have been tested against exact solutions of the Boltzmann equation for  systems subject to Bjorken \cite{Florkowski:2013lza,Florkowski:2013lya,Florkowski:2014sfa,Strickland:2017kux,Florkowski:2017ovw,Strickland:2018ayk} and Gubser flows \cite{Denicol:2014xca,Nopoush:2014qba,Denicol:2014xca,Martinez:2017ibh,Behtash:2017wqg}.  In all cases, it was found that aHydro provided the best approximation to the exact solutions for both hydrodynamic and non-hydrodynamic moments of the distribution function \cite{Strickland:2018ayk}.

This provided motivation to compare the aHydro framework with experimental results.  Despite the success of these early comparisons, in all phenomenological applications of aHydro to date, leading-order aHydro codes have been implemented using an anisotropy tensor which possesses only diagonal (elliptical) anisotropies (see Ref.~\cite{Alqahtani:2017mhy} for a recent review).  This was done mainly because of the difficulty of efficiently evaluating the necessary moment integrals in the presence of off-diagonal anisotropies $\xi^{ij}$ with $i \neq j$.  However, to be complete, one must also include the possibility of off-diagonal leading-order anisotropies.  Near equilibrium, this is equivalent to including off-diagonal components in the LRF shear viscous tensor $\pi^{ij}$.  

In this paper, we present a technique that can be used to efficiently include non-vanishing $\xi^{ij}$.  This is done by a change of variables in the generic moment integrals which diagonalizes the anisotropy tensor.  Once cast into diagonal form, a previously developed technique for the efficient application of diagonal moment integrals can be used to compute the necessary off-diagonal moment integrals (see Appendix B of Ref.~\cite{Alqahtani:2017tnq}).  We present the general method of diagonalization and provide some concrete examples for the application to aHydro frameworks which use the so-called anisotropic-pressure- or Tinti-matching \cite{Tinti:2015xwa,Molnar:2016vvu}.

\section*{Conventions and notation}

The Minkowski metric tensor is taken to be ``mostly minus'', i.e. $g^{\mu\nu}={\rm diag}(+,-,-,-)$. The vector $u^\mu$ is the flow velocity which satisfies normalization condition $u_\mu u^\mu=1$. The transverse projection operator $\Delta^{\mu\nu}\equiv g^{\mu\nu}{-}u^\mu u^\nu$ is used to project four-vectors and/or tensors into the space orthogonal to $u^\mu$. Parentheses and square brackets on indices denote symmetrization and anti-symmetrization, respectively, i.e. $A^{(\mu\nu)}\equiv\frac{1}{2}\left(A^{\mu\nu}{+}A^{\nu\mu}\right)$  and $A^{[\mu\nu]}\equiv\frac{1}{2}\left(A^{\mu\nu}{-}A^{\nu\mu}\right)$. Angle brackets on indices indicate projection with a four-index transverse projector, \mbox{$A^{\langle \mu \nu\rangle}\equiv\Delta^{\mu\nu}_{\alpha\beta}A^{\alpha\beta}$}, where $\Delta^{\mu\nu}_{\alpha\beta}\equiv\Delta^{(\mu}_\alpha\Delta^{\nu)}_\beta-\Delta^{\mu\nu}\Delta_{\alpha\beta}/3$ projects out the traceless and $u^\mu$-transverse components of a rank-two tensor.  The Lorentz-invariant momentum-space integration measure is indicated as $dP=\tilde{N}d^3{\bf p}/(p\cdot u)$, with $\tilde{N}=N_{\rm dof}/(2\pi)^3$ where $N_{\rm dof}$ is the number of degrees of freedom. 

In order to write the equations of motion in a manifestly Lorentz-covariant manner it is useful to introduce the LRF basis vectors, as $u^\mu_{\rm LRF}=(1,{\bf 0})$ and $X^\mu_{i,{\rm LRF}}=(0,{\boldsymbol \delta}^\mu_i)$ with $i \in \{1,2,3\}$. By applying a sequence of Lorentz transformations, one can construct the lab frame basis vectors, i.e. $u^\mu$ and $X^\mu_i$ with $i \in \{1,2,3\}$, where the dynamical equations are solved and particle spectra are computed \cite{Ryblewski:2010ch,Martinez:2012tu}. It is also useful to define the transverse projection operator in terms of the space-like basis vectors, i.e. $\Delta^{\mu\nu}=-\sum_i X^\mu_i X^\nu_i$. Finally, note that the Latin indices sum over space-like indices (components of three-vectors) and Greek indices sum over components of four-vectors.

\section{Leading-order anisotropic hydrodynamics}
\label{sec:anisoDist}

In leading-order aHydro, the one-particle distribution function is parametrized by an anisotropy tensor which results in the deformation of the argument of an isotropic distribution function into an anisotropic one \cite{Martinez:2012tu,Nopoush:2014pfa}
\ba 
f_a(x,p)=f_{\rm iso}\bigg(\frac{1}{\lambda}\sqrt{p_\mu\Xi^{\mu\nu}p_\nu}\bigg)\,,
\label{eq:anisodist1}
\ea
%
where $\lambda$ has dimensions of energy and can be identified with temperature only in the isotropic equilibrium limit.  In practice, $f_{\rm iso}$ can be a Bose-Einstein, Fermi-Dirac, or Maxwell-Boltzmann distribution depending on particle statistics and/or energy.  In the non-conformal (massive) case, the rank-2 tensor $\Xi^{\mu\nu}$ specifying the shape of the distribution in momentum space is defined as  \cite{Martinez:2012tu,Nopoush:2014pfa}
\ba 
\Xi^{\mu\nu}=u^\mu u^\nu +\xi^{\mu\nu}-\Phi \Delta^{\mu\nu}\,,
\label{eq:xi}
\ea
%
where $\xi^{\mu\nu}$ denotes a symmetric traceless anisotropy tensor, i.e. $\xi_x+\xi_y+\xi_z=0$ in the LRF.  

The quantities $\lambda$, $u^\mu$, and $\xi^{\mu\nu}$ are spacetime fields which satisfy the following identities
\ba
u^\mu u_\mu &=& 1  \, ,  \\
{\xi^{\mu}}_\mu &=& 0  \, , \\
u_\mu \xi^{\mu\nu} &=& 0 \, .
\ea
%
The third condition above, indicating orthogonality of $\xi^{\mu\nu}$ to $u^\mu$ which implies that, in the LRF, $\xi^{\mu\nu}$ obeys the following conditions
\ba
\xi^{00}=\xi^{0i}=\xi^{i0}=0\,.
\ea
%
Working in the LRF, this allows us to focus on the non-trivial space-like components of $\xi^{\mu\nu}$ as $\bm{\xi}$, which is a $3\times 3$ matrix.
The argument of distribution function subject to the mass-shell condition can be simplified as
\ba
p\cdot\Xi \!\cdot p={\bf p}\cdot \kappa \cdot {\bf p}+m^2\,,
\ea
%
which gives
\ba
f_a(x,p)=f_{\rm iso}\left(\frac{1}{\lambda}\sqrt{{\bf p}\cdot \kappa\cdot {\bf p}+m^2}\right)\,,
\label{eq:distfunc}
\ea
where 
\ba
\kappa \equiv \bm{I}(1+\Phi)+\bm{\xi}\,,
\ea
%
with $\bm{I}$ being a $3\times 3$ identity matrix. 

If $\bm{\xi}$ is diagonal, i.e.
\be
\bm{\xi}={\rm diag}(\xi_x,\xi_y,\xi_z)\,,
\label{eq:xi-lower}
\ee
which implies the ellipsoidal distribution, the $\kappa$ matrix is automatically diagonal, i.e.
\mbox{$
\kappa={\rm diag}(1/\alpha_x^2,1/\alpha_y^2,1/\alpha_z^2)
$}
with $\alpha_i=(1+\xi_i+\Phi)^{-1/2}$ \cite{Nopoush:2014pfa}.
For a non-ellipsoidal distribution function, generalizing $\bm{\xi}$ to include off-diagonal components, one has
\be
\kappa=
\begin{pmatrix}
1/\alpha_x^2&\xi_{xy}&\xi_{xz}\\\xi_{xy}&1/\alpha_y^2&\xi_{yz}\\\xi_{xz}&\xi_{yz}&1/\alpha_z^2
\end{pmatrix} \,.
\label{eq:kappa}
\ee
%
Note that, in a general frame one has $\xi^{\mu\nu} = \kappa_{ij} X^\mu_i X^\nu_j$ where the summation over $i$ and $j$ is implied.

\section{Diagonalization}

Calculating the bulk variables in aHydro requires computing momentum-space moments of the distribution function. However, the distribution function in Eq.~(\ref{eq:distfunc}) is a complicated function of momentum and there is no way to perform the integrals analytically except in some special cases. 
In this section, we introduce an algebraic method to diagonalize the $\kappa$ matrix so that we can reduce the computation of moment-integrals including off-diagonal anisotropies to a linear combination of diagonal momentum-space moment integrals.

 For any $N\times N$ real and symmetric matrix $\kappa$ there exists a unitary matrix $A$ such that
 \ba
\kappa=A\,\kappa_D A^\dagger\,,
\label{eq:kapa}
\ea
where $A$ is constructed such that its columns are the eigenvectors of $\kappa$.
The combination ${\bf p}\cdot\kappa\cdot {\bf p}$ can be written as
\ba
{\bf p}\cdot\kappa\cdot {\bf p}= {\bf p}^T \kappa\,{\bf p}=\Big[{\bf p}^T A\Big]\Big[ A^\dagger \kappa A\Big]\Big[A^\dagger {\bf p}\Big]=\tilde{\bf p}^T\, \kappa_D\, \tilde{{\bf p}}=\tilde{\bf p}\cdot\kappa_D\cdot \tilde{\bf p}\,,
\label{eq:kappa-trans}
\ea
%
with $\tilde{\bf p}\equiv A^\dagger{\bf p}$. 
By definition we have 
\ba
{\bf p}=A\tilde{\bf p} \quad\Rightarrow\quad {p}_i=\sum_j A_{ij}\tilde{p}_j\,.
\ea
%
For example
\ba
p_i=\sum_j A_{ij} \tilde{p}_j=\sum_j v_i^{(j)}\tilde{p}_j, \label{eq:p-transform}
\ea
%
where the vector $v^{(i)}=(v_x^{(i)},v_y^{(i)},v_z^{(i)})$ is the $i^{\rm th}$ eigenvector of $\kappa$. Therefore, we have two frames, i.e. the original frame and the rotated frame, where the components of the momentum vector are ${\bf p}_i$ and $\tilde{\bf p}_i$, respectively. The $\kappa$ matrix in the original frame is defined in Eq.~(\ref{eq:kappa}) and, in the rotated frame, is defined as $\kappa_D\equiv {\rm diag}(1/\tilde{\alpha}_i^2)$. These two frames are connected by rotations through a set of Euler angles. Note that the Jacobian for transforming between two frames is unity.  

It is obvious that the length of $\bf p$ is invariant under this coordinate transformation. Accordingly, as expected, $E$ is the same in both coordinate systems
\ba
 E=\sqrt{{\bf p}^2+m^2}=\sqrt{\tilde{\bf p}^2+m^2} = \tilde{E}\,.
 \label{eq:E}
\ea
%
Using Eq.~(\ref{eq:p-transform}), one can simplify the general anisotropic distribution function to the anisotropic distribution function (\ref{eq:distfunc})  with diagonal anisotropy tensor (\ref{eq:kappa}) in the rotated frame
\ba
f_a(x,p)=f_{\rm iso}\left(\frac{1}{\lambda}\sqrt{{\bf p}\cdot \kappa\cdot {\bf p}+m^2}\right)=f_{\rm iso}\left(\frac{1}{\lambda}\sqrt{\tilde{\bf p}\cdot\kappa_D\cdot \tilde{\bf p}+m^2}\right)\equiv f^D_a(x,\tilde{p}) \,.
\label{eq:distfuncD}
\ea

\section{The energy-momentum tensor}

We begin by demonstrating how this method can be used to efficiently evaluate the components of the energy-momentum tensor including off-diagonal anisotropies.  In the general case, we have six independent anisotropy parameters ($\alpha_x$, $\alpha_y$, $\alpha_z$, $\xi_{xy}$, $\xi_{xz}$, and $\xi_{yz}$), one momentum-scale parameter ($\lambda$), and the three independent components of the fluid four-velocity ($u^i$), resulting in ten space-time fields for which we must obtain equations of motion.  In the LRF, the non-vanishing components of the energy-momentum tensor are
\ba
 T^{00} &=&{\cal E}= \int dP \, E^2 \, f_a(x,p) \, , \\
T^{ij} &=&\int dP \, p^i p^j  \, f_a (x,p) \, .
\ea
%
Using the techniques introduced in the previous section, one finds
\ba
{\cal E}&=&\int dP \, E^2 \, f_a (x,p)  \nonumber \\ 
&=& \tilde{N} \int d^3\tilde{\bf p} \, \sqrt{\tilde{\bf p}^2+m^2} \, f^D_a(x,\tilde{p})
=\tilde{\alpha}\lambda^4 Q_3(\tilde{\alpha}^2_x,\tilde{\alpha}^2_y,\tilde{\alpha}^2_z,\hat{m}) \, , 
\label{eq:E-final}
\ea
%
and
\ba
T^{ij} &=&\int dP \, p^i p^j \, f_a (x,p) \nonumber \\
&=& \tilde{N} \int \frac{d^3\tilde{\bf p}}{\sqrt{\tilde{\bf p}^2+m^2}}\, f^D_a(x,\tilde{p}) \sum_{k,l=1}^3 v_i^{(k)}v_j^{(l)}\tilde{p}^k\tilde{p}^l= \tilde{\alpha} \lambda^4 \sum_{k=1}^3 v_i^{(k)} v_j^{(k)} \tilde{\alpha}_k^2 \, Q_3^{k} (\tilde{\alpha}^2_x,\tilde{\alpha}^2_y,\tilde{\alpha}^2_z,\hat{m})\,. \quad 
\label{eq:tijfinal}
\ea
%
The $Q$-functions appearing above only depend on the diagonal anisotropies $\tilde{\boldsymbol \alpha}$ and are defined in Appendix \ref{app:q}.  The scaled mass variable is defined as $\hat{m} \equiv m/\lambda$ and we have introduced a compact notation as $\tilde{\alpha} \equiv \tilde{\alpha}_x  \tilde{\alpha}_y  \tilde{\alpha}_z$. Based on the symmetry of $T^{ij}$ under exchanging the indices, out of 9 possible values there are only 6 unique terms that must be calculated.
 Note that for the diagonal terms (pressures) one obtains
\be
{\cal P}_i = T^{ii} = \tilde{\alpha} \lambda^4 \sum_{k=1}^3 \left[v_i^{(k)}\right]^2 \tilde{\alpha}_k^2\, Q_3^k (\tilde{\alpha}^2_x,\tilde{\alpha}^2_y,\tilde{\alpha}^2_z,\hat{m}) \, .
\ee
In all cases above, we have reduced the problem to computing $Q$-functions with only diagonal anisotropies.  The diagonal anisotropy tensor integrals can be well-approximated by Taylor expanding to high-order around an isotropic point, e.g. $\tilde{\boldsymbol \alpha}_{\rm iso} = (\alpha_0,\alpha_0,\alpha_0)$.  At each order in this expansion the required integrals can be performed analytically.  In order to cover the space using truncated Taylor expansions, one can utilize multiple expansion points which are then pieced together to accurately span the range of diagonal anisotropies which are generated in typical simulations.  Using modern computerized algebra systems one can extend the Taylor expansion expressions described above to high order.  In practice, phenomenological codes have used $12^{\rm th}$ order truncations in $\tilde{\boldsymbol \delta} = \tilde{\boldsymbol \alpha} -\tilde{\boldsymbol \alpha}_{\rm iso}$ (see Appendix B of Ref.~\cite{Alqahtani:2017tnq}).

\section{Dynamical equations - Anisotropic pressure matching}

To further demonstrate the utility of this method, we now consider equations for the viscous tensor obtained by anisotropic pressure matching \cite{Tinti:2015xwa}.  In relaxation-time approximation (RTA) the dynamical equations for the shear and bulk viscous corrections based on anisotropic matching are 
\begin{eqnarray}
\partial_\mu T^{\mu\nu}=0\,, \label{emcons} \\
  D_u\pi^{\langle\mu\nu\rangle} + \frac{1}{\taueq} \pi^{\mu\nu} &=&  -  \left( \sigma_{\rho \sigma} +\frac{1}{3} \, \theta \, \Delta_{\rho\sigma}  \right) \int dP \,\frac{ p^{\langle\mu} p^{\nu\rangle} p^\rho p^\sigma  \,  f_a}{(p\cdot u)^2} - 2 \, \pi_\alpha^{<\mu}\sigma^{\nu>\alpha} \nonumber \\ 
  && + 2\, \pres \, \sigma^{\mu\nu} -\frac{5}{3}\, \theta \, \pi^{\mu\nu} + 2 \, \pi_\alpha^{<\mu}\omega^{\nu>\alpha}, \label{shear_RTA_LO_1} \\  
D_u\pres + \frac{1}{\taueq}\left( \pres-\preseq \right) &=&  \frac{1}{3} \left(\sigma_{\rho \sigma} +\frac{1}{3} \, \theta \, \Delta_{\rho\sigma} \right)  \int dP \,\frac{ (p\cdot \Delta \cdot p) p^\rho p^\sigma  \,  f_a}{(p\cdot u)^2} 
 +\frac{2}{3} \, \pi_{\mu\nu}\sigma^{\mu\nu} - \frac{5}{3} \, \pres \,  \theta \, .\qquad
\label{bulk_RTA_LO_1}
\end{eqnarray}
%
In the above relations, $f_a$ is the general distribution function defined at Eq.~(\ref{eq:anisodist1}). The tensor $\pi^{\mu\nu}$ is the shear tensor, which is traceless and orthogonal to flow velocity $u^\mu$.  In the relations above one has ${\cal P}={\cal P}_{\rm eq}+\Pi$ with ${\cal P}_{\rm eq}  $ being the LRF equilibrium pressure which can be obtained by evaluating any component of Eq.~(22) with $\kappa$ equal to an identity matrix and $\lambda$ set to the local effective temperature $T$.  The other symbols appearing in Eqs.~\eqref{shear_RTA_LO_1} and \eqref{bulk_RTA_LO_1} above are defined as
\be
\begin{aligned}
D_u &= u^\mu \partial_\mu \, , \\
D_i&=X^\mu_i \partial_\mu\,,\\
\theta &=\nabla_\mu u^\mu\,,
\end{aligned}
\hspace{2.5cm}
\begin{aligned}
\nabla^\mu &=\Delta^{\mu\nu}\partial_\nu\,,\\
\omega^{\mu\nu}&=(\nabla^\mu u^\nu-\nabla^\nu u^\mu)/2\,,\\
\sigma^{\mu\nu}&=\Delta^{\mu\nu}_{\alpha\beta}\partial^\alpha u^\beta\,.
\end{aligned}
\label{eq:identities}
\ee
The equations (\ref{bulk_RTA_LO_1}) represent a set of ten dynamical equations for the ten independent macroscopic variables of the system. Microscopically, one has three components of flow velocity $u^i$, six independent anisotropy parameters, and the temperature-like scale $\lambda$,  resulting in ten dynamical microscopic variables. Correspondingly, when coding up these equations, one can choose between using macroscopic or microscopic variables.  In addition, if using the macroscopic variables, one can evolve the ten independent components of (symmetric) energy-momentum tensor $T^{\mu\nu}$ or one can use the standard decomposition \cite{McNelis:2018jho,Tinti:2015xwa}
\ba 
T^{\mu\nu}=T^{\mu\nu}_{\rm eq}+\pi^{\mu\nu}+\Pi \Delta^{\mu\nu}\,,
\label{eq:T-exp}
\ea
which has as dynamical variables ${\cal E}$, three components of flow velocity $u^i$, five independent components of shear tensor $\pi^{\mu\nu}$, and the bulk viscous correction $\Pi$, again added up to ten. 

In practice, it is preferable to evolve the macroscopic (thermodynamics) variables, since modern flux-conserving algorithms are better suited to these equations than the microscopic ones. However, this procedure is non-trivial because, although the above equations evolve macroscopic variables, they explicitly contain microscopic ones as well, e.g. the distribution function $f_a$ appearing in Eq.~(\ref{eq:distfunc}). Therefore, in order to close the system of equations one must update the microscopic variables in parallel to the macroscopic ones during the evolution. Roughly speaking, the procedure is as follows: The equation $\partial_\mu T^{\mu\nu}=0$ provides the evolution of ${\cal E}$ and $u^i$. The other equations evolve the components of the shear tensor. Using these, one can construct the full $T^{\mu\nu}$ using (\ref{eq:T-exp}). Once the lab frame $T^{\mu\nu}$ is evolved forward one time step, the updated microscopic variables can be obtained by boosting to the LRF and solving a set of seven coupled matching equations which match $T^{00}_{\rm LRF}$, and six components of upper diagonal space-like block of $T_{\rm LRF}^{\mu\nu}$ to their microscopic definitions as a function of $\alpha_i$, $\xi_{ij}$, and $\lambda$, i.e. Eqs.~(\ref{eq:E-final}) and (\ref{eq:tijfinal}).

In order to further develop the necessary formalism, one must expand and simplify the dynamical equations (\ref{bulk_RTA_LO_1}) for the case of a non-ellipsoidal anisotropic distribution function. Note that we will expand the equations in the lab frame, where the dynamical equations are solved. However, whenever a scalar quantity is obtained, we have the freedom to choose a covariant Lorentz frame, e.g. local reference frame, where the calculation is simpler. 

There are two terms in Eqs.~\eqref{shear_RTA_LO_1} and \eqref{bulk_RTA_LO_1} needing detailed expansion. The first one is
\ba
\left(\sigma_{\rho\sigma}+\frac{1}{3}\theta\Delta_{\rho\sigma}\right)p^\rho p^\sigma= p^\rho p^\sigma \nabla_\rho u_\sigma=p^\sigma({\bf p}\cdot {\bf D})u_\sigma\,,
\ea
%
where ${\bf D}$ is defined in (\ref{eq:identities}). We also have 
\ba
p^{\langle\mu}p^{\nu\rangle} =p^\alpha p^\beta \Delta_\alpha^\mu  \Delta_\beta^\nu+\frac{1}{3}\Delta^{\mu\nu} {\bf p}^2=p^i p^j X_i^\mu X_j^\nu+\frac{1}{3}\Delta^{\mu\nu} {\bf p}^2\,,
\ea
%
where the Einstein summation convention for repeated spatial indices is applied. The very last step is performed in order to make the dependence of components of momentum 3-vector explicit, which is useful in evaluating the integrals necessary. 

The other term is
\ba
p\cdot \Delta\cdot p=-{\bf p}^2\,.
\ea
%
Using the above relations, one can expand the following integrals
\ba
 &&-\bigg(\sigma_{\rho\sigma}+\frac{1}{3}\theta\Delta_{\rho\sigma}\bigg)\int \frac{dP}{E^2}p^\rho p^\sigma p^{\langle \mu}p^{\nu\rangle}f_a\nonumber \\ 
 &&\hspace{1.7cm}=-\int \frac{dP}{E^2}f_a\, p^i p^jX_i^\mu X_j^\nu p^\sigma({\bf p}\cdot {\bf D}) u_\sigma-\frac{\Delta^{\mu\nu}}{3}\!\int \frac{dP}{E^2}f_a \,{\bf p}^2 p^\sigma({\bf p}\cdot {\bf D}) u_\sigma\nonumber \\
 &&\hspace{1.7cm}=\int \frac{dP}{E^2}f_a \,p^i p^jX_i^\mu X_j^\nu p^l({\bf p}\cdot {\bf D}) u_l+\frac{\Delta^{\mu\nu}}{3}\!\int \frac{dP}{E^2}f_a \, {\bf p}^2 p^l({\bf p}\cdot {\bf D}) u_l\nonumber \\
&&\hspace{1.7cm}= \left[{\cal F}^{ijkl}\,X_i^\mu X_j^\nu  +\frac{\Delta^{\mu\nu}}{3}{\cal F}^{iikl} \right] D_k u_l\,,
 \label{eq:ex1}
 \ea
where the four-index function introduced above is defined as 
\ba
{\cal F}^{ijkl}&\equiv&\int \frac{dP}{E^2} \, p^i p^j p^k p^l f_a(x,p)\,.
\label{eq:F2}
\ea
Note that for Eq.~(\ref{eq:ex1}) to be non-vanishing one must have an even number of spatial momenta with matching indices, appearing in ${\cal F}^{ijkl}$. To see this, consider the integral above containing an odd number of spatial momenta. Using the map  (\ref{eq:p-transform}) it will contain an odd number of $\tilde{p_i}$ even in the rotated frame and the rest of the integrand will be an even function of the momenta. Therefore, the integral will vanish by symmetry in this case.  This suggests
 that in the third line of the equation (\ref{eq:ex1}) defined above, $p^\sigma u_\sigma \rightarrow -p^i u_i$.

Similarly, the non-trivial term appearing in the bulk viscous equation of motion \eqref{bulk_RTA_LO_1} is
\ba
\frac{1}{3}\left(\sigma_{\rho\sigma}+\frac{1}{3}\theta\Delta_{\rho\sigma}\right)\int \frac{dP}{E^2}f_a \,p^\rho p^\sigma (p\cdot\Delta\cdot p)
&=&-\frac{1}{3}\int \frac{dP}{E^2} \,f_a\, {\bf p}^2p^\beta({\bf p}\cdot{\bf D}) u_\beta \nonumber\\
&=&\frac{1}{3}\int \frac{dP}{E^2} \,f_a\, {\bf p}^2p^l({\bf p}\cdot{\bf D}) u_l \nonumber\\
&=&\frac{1}{3}{\cal F}^{iikl}\,\partial_k u_l \, . 
 \label{eq:ex2}
\ea
%
To complete the simplification of the non-trivial terms in (\ref{eq:ex1}) and (\ref{eq:ex2}), we now consider the ${\cal F}$ function.  Using similar techniques as used for the $T^{ij}$, one obtains
\ba
{\cal F}^{ijkl} &=&\tilde{N}\int \frac{d^3{\bf p}}{E^3} p^i p^jp^k p^l\,f_a(x,{\bf p})=\tilde{N}\int \frac{d^3 \tilde{\bf p}}{E^3} f_a^D(x,{\tilde{\bf p}}) \, {\cal P}_{mn}\Big[v_i^{(m)} v_j^{(m)} v_l^{(n)} v_k^{(n)}\Big]\tilde{p}_m^2 \,\tilde{p}_n^2\,,\nonumber\\
&=& \tilde{\alpha}\lambda^4  {\cal P}_{mn}\Big[v_i^{(m)} v_j^{(m)} v_l^{(n)} v_k^{(n)}\Big]  \tilde{\alpha}_m^2\tilde{\alpha}_n^2\,Q_3^{mn}(\tilde{\alpha}^2_x,\tilde{\alpha}^2_y,\tilde{\alpha}^2_z,\hat{m})\,.
\ea
%
The operator ${\cal P}_{mn}$ introduced above is the permutation operator which sums over all possible permutations of $m$ and $n$ in the operand (including repeated ones).  Based on the symmetry of ${\cal F}^{ijkl}$ under exchanging the indices, out of 81 possible values there are only 15 unique terms that must be calculated.  The function $Q^{mn}$ introduced above is defined in Appendix \ref{app:q}.

\section{Discussion and summary}

As we demonstrated in the previous two sections, one can reduce the problem of evaluating complicated off-diagonal anisotropy moment integrals to a sum of diagonal anisotropy integrals.  In practice, one can use Eqs.~\eqref{emcons}, \eqref{shear_RTA_LO_1}, and \eqref{bulk_RTA_LO_1} to evolve the energy-momentum tensor, shear viscous tensor, and the bulk viscous correction, respectively.  Given an initial condition specified in terms of all anisotropies and the momentum scale, $\lambda$, one can construct the full energy-momentum tensor at the initial time.  One can then evolve the coupled partial differential equations \eqref{emcons}, \eqref{shear_RTA_LO_1}, and \eqref{bulk_RTA_LO_1} forward in time by one infinitesimal step making use of the methods explained in the previous section to evaluate the non-trivial integrals involving $f_a$ in Eqs.~\eqref{shear_RTA_LO_1} and \eqref{bulk_RTA_LO_1}.  

Once the update is complete, one can solve a set of seven non-linear equations to extract the updated LRF anisotropies and scale parameter.  These can then be used to compute the non-trivial integrals involving $f_a$ in the next time step.  Repeating this procedure, one can evolve all dynamical fields using Eqs.~\eqref{emcons}, \eqref{shear_RTA_LO_1}, and \eqref{bulk_RTA_LO_1}.  Critical to accomplishing this is the efficient evaluation of the integrals involving $f_a$ in Eqs.~\eqref{shear_RTA_LO_1} and \eqref{bulk_RTA_LO_1} and the subsequent extraction of the local anisotropy tensor from the full energy-momentum tensor.  The diagonalization method described in the previous two sections solves this problem by removing the bottleneck of evaluating complicated three dimensional integrals on demand.

\section{Conclusions}

In this paper we presented a method for efficiently including the effects of off-diagonal local rest frame momentum anisotropies in leading-order anisotropic hydrodynamics.  The method relies on diagonalization of the space-like block of the anisotropy tensor and allows one to reduce the necessary moments of the distribution function in the off-diagonal case to a linear combination of diagonal-anisotropy integrals.  Once reduced to diagonal-anisotropy integrals, the results can be computed efficiently using techniques described previously in the literature \cite{Alqahtani:2017tnq}.  We presented a general framework for how to accomplish this and provided examples for off-diagonal anisotropy moments entering into the energy-momentum tensor and viscous update equations which emerge when performing anisotropic pressure matching \cite{Tinti:2015xwa}.  With this method in hand one can implement a leading-order anisotropic hydrodynamics code that takes into account off-diagonal anisotropies non-perturbatively.  Additionally, since the equations are formulated at the level of the energy-momentum tensor and shear viscous tensor, this more easily allows for the use of advanced numerical techniques for solving the necessary partial differential equations (see e.g. \cite{Bazow:2016yra}).

\acknowledgments

We thank L. Tinti for early conversations about how to handle off-diagonal anisotropies.  M. Strickland was supported by the U.S. Department of Energy, Office of Science, Office of Nuclear Physics under Award No. DE-SC0013470.

\appendix

\section{Q-functions}
\label{app:q}

The Q-functions used in expanding the equations are defined as follows
\ba
Q_3(\alpha^2_x,\alpha^2_y,\alpha^2_z,\hat{m}) &=& \tilde{N}\int d^3{\bf p} \, \sqrt{\sum_k \alpha_k^2\, p_k^2+\hat{m}^2} \, f_{\rm iso}\left(\sqrt{{\bf p}^2+\hat{m}^2}\right)\,, \label{eq:Q3} \\
Q_3^i(\alpha^2_x,\alpha^2_y,\alpha^2_z,\hat{m}) &=& \tilde{N} \int d^3{\bf p} \, \frac{p_i^2}{\sqrt{\sum_k\alpha_k^2\, p_k^2+\hat{m}^2}}\, f_{\rm iso}\left(\sqrt{{\bf p}^2+\hat{m}^2}\right)\,,\label{eq:Q3i}  \\
Q_3^{ij}(\alpha^2_x,\alpha^2_y,\alpha^2_z,\hat{m}) &=&\tilde{N}  \int d^3{\bf p} \, \frac{p_i^2 p_j^2}{\Big(\sum_k \alpha_k^2\, p_k^2 +\hat{m}^2\Big)^{3/2}}\, f_{\rm iso}\left(\sqrt{{\bf p}^2+\hat{m}^2}\right) \,.
\label{eq:Q3ij}
\ea
We note that the functions above functions are related, e.g.
\ba
 Q_3^i&=&2\frac{\partial Q_3}{\partial \alpha_i^2}\,,\\
 Q_3^{ij}&=&-2\frac{\partial Q^i_3}{\partial \alpha_j^2}=-4\frac{\partial^2 Q_3}{\partial \alpha_i^2\partial\alpha_j^2}\,.
\ea
%
This fact allows us to reduce the number of underlying $Q$-functions that have to be computed to the ``master function'' $Q_3$.

\bibliography{OD-ahydro}

\end{document}